\documentclass[aps,onecolumn,preprintnumbers,groupedaddress,nofootinbib]{revtex4}
\usepackage{graphics}
\usepackage{epsfig}

\usepackage{graphicx}
\IfFileExists{srcltx.sty}{\usepackage[active]{srcltx}}

\begin{document}
\title{Properties of  Q-ball Dark Matter: \\ Moving Away from Flat Directions}   %%% Fill in title
\author{Ian M. Shoemaker}   %%% Fill in author names
\affiliation{Department of Physics and Astronomy, University of California, Los 
Angeles, CA 90095-1547, USA }
\preprint{UCLA/09/TEP/52}

\begin{abstract} %%% Abstract to run on from here.
    \centering
    \begin{minipage}{0.8\textwidth}
Gauge-mediated models of supersymmetry-breaking imply that stable $Q$-balls can form in the early universe and act as dark matter. All  stable $Q$-balls in the MSSM are associated with one or more flat directions. We show that while $Q$-balls are produced from the fragmentation of a flat direction condensate, they quickly evolve to a ground state that is slightly away from this flat direction. In this process a $(B+L)$-ball can become electrically charged. This is a novel form of charge acquisition with important ramifications for the experimental search for $Q$-ball dark matter. 
\end{minipage}
\end{abstract}

\maketitle
%%%%
\section{Introduction}
%%%%

A field theory with one or more scalar fields that conserves a global $U(1)$ symmetry often allows for the existence of non-topological solitons, dubbed $Q$-balls \cite{Friedberg:1976me, Coleman:1985ki, Lee:1991ax}. Given a constant amount of global charge $Q$, the $Q$-ball is the state that minimizes the energy.  It has been shown \cite{Kusenko:1997ad} that all supersymmetric generalizations of the Standard Model admit $Q$-ball solutions, where the global charge is provided by the baryon and lepton numbers. However, while the $Q$-ball is by construction stable against decay into scalars it may not necessarily be stable against decay into fermions \cite{Cohen:1986ct}.  In the gauge mediated scenario of supersymmetry breaking however, $Q$-balls are entirely stable \cite{Kusenko:1997si}. These $Q$-balls are formed in the early universe from the fragmentation of the Affleck-Dine condensate \cite{Affleck:1984fy, Enqvist:2003gh, Dine:2003ax}. In addition to being attractive dark matter candidates, their partial evaporation can also naturally explain the origin of the ratio $\Omega_{b} / \Omega_{DM} \sim 1$ \cite{Laine:1998rg}. 

The cosmological implications of $Q$-balls have been extensively studied \cite{Dine:2003ax,Enqvist:2003gh}. We briefly summarize some of the recent activity in this area.  It has been demonstrated that the out-of-equilibrium decay of unstable Q-balls into LSPs can naturally explain the PAMELA and ATIC data \cite{McDonald:2009cc}. In addition, the fragmentation of the primordial condensate into $Q$-balls can generate an observable spectrum of gravitational waves \cite{Kusenko:2008zm, Kusenko:2009cv}. Moreover, both the quantum and absolute stability of $Q$-balls in supersymmetric models has been given a thorough treatment \cite{Copeland:2009as}, and the existence of $Q$-balls with nonzero angular momentum demonstrated in \cite{Campanelli:2009su}. It has also been found that the terrestrial passage of dark matter $Q$-balls can lead to an observable flux of neutrinos with a unique zenith angle dependence \cite{Kusenko:2009iz}.

 In the Minimal Supersymmetry Standard Model (MSSM) the formation of $Q$-balls occurs along one of the many flat directions. Composed of some combination of squark, slepton, and Higgs fields these are directions in scalar field space along which the classical scalar potential vanishes \cite{Gherghetta:1995dv}.

The mechanism by which supersymmetry is communicated to the visible sector dictates the type of $Q$-ball that can form.  If supersymmetry breaking is only communicated only via gravity, then $Q$-balls are unstable \cite{Enqvist:2003gh}. However if supersymmetry is broken via low-energy gauge-mediation then the $Q$-ball can be entirely stable. In this scenario the mass of the $Q$-ball is
\begin{equation}
M_{Q} \sim M_{s} Q_{B}^{3/4}, 
\label{mass}
\end{equation}
where $M_{s}$ is the scale of supersymmetry breaking, and $Q$ is the baryon number of the $Q$-ball. As long as the mass per charge inside the $Q$-ball is less than the mass per charge outside ($m_{p} \sim 1~\textsl{GeV}$), decay into baryons will be kinematically forbidden. Moreover, since the mass per charge inside the $Q$-ball is a monotonically decreasing function of $Q_{B}$ we are always guaranteed a stable solution for sufficiently large $Q_{B}$: 
\begin{equation} 
\omega \equiv \frac{M_{Q}}{Q} \sim \frac{M_{s}}{Q^{1/4}} \ll 1~ \textsl{GeV}. \end{equation}
Thus with $M_{s} \sim \textsl{TeV}$ we need $Q_{B} \gg 10^{12}$ to ensure stability.  Although this is an extremely large baryon number, an attractive production mechanism for large $Q$-balls is offered by the Affleck-Dine Mechanism of baryogenesis.  It has been shown that in this scenario such huge $Q$-balls are quite natural, and that these $Q$-balls are attractive dark matter candidates \cite{Kusenko:1997si}

It has also been shown that another type of stable $Q$-ball exists in the gauge-mediated scenario \cite{Kasuya:2000sc}. When the field amplitude inside the $Q$-ball becomes sufficiently large the effects of gravity-mediation become important and both sources of supersymmetry-breaking must be taken into account. In this case the $Q$-ball has a mass $M_{Q} \sim m_{3/2} Q_{B}$, which makes the $Q$-ball stable since the gravitino mass is generally quite small ($m_{3/2} \ll 1~ \textsl{GeV}$) in models of gauge-mediation. This stability allows these $Q$-balls to be dark matter candidates as well. In the present work however we restrict ourselves to sufficiently small VEV such that gravity effects are negligible. 

It has been argued that $(B+L)$-balls always eventually transform into pure $B$-balls because the leptonic component can always radiate away in the form of neutrinos \cite{Kawasaki:2004th}.  This is true if one assumes that the $Q$-ball does not deviate from the flat direction as it evolves. We show \cite{Shoemaker:2008gs} however that in the gauge-mediated case the true ground state violates the condition of $D$-flatness.  Although the baryon number is constrained for overall stability, the lepton number can deviate slightly from the flat direction value as the $Q$-ball evolves to the true ground state. This is a new mechanism for the acquisition of electric charge for an otherwise neutral $Q$-ball.  The experimental detection strategies differ greatly between neutral and charged $Q$-balls \cite{Ambrosio:1999gj, Arafune:2000yv, Takenaga:2006nr}. The true ground state configuration is therefore crucial for the prospects of experimental detection. 

The remainder of the paper is organized as follows. Section 2 reviews the properties of flat-direction $Q$-balls in the MSSM. Section 3 examines ground state $Q$-ball solutions as the flat-direction $Q$-ball decays. In section 4 we apply our results to the $QdL$ flat direction in the MSSM. Having established the existence of non-flat direction states in the MSSM, we examine the time evolution from FD to ground state confiurgations in Section 5. We summarize the experimental limits on these new $Q$-balls in Section 6, and conclude in Section 7.

%%%
\section{Flat Direction \boldmath $Q$-balls} 
\label{sec1}
%%%

Although a flat direction can generally be parameterized by a single scalar degree of freedom, the gauge eigenstates and the flat direction parameter may be non-trivial \cite{Enqvist:2003pb}. In the case that a single degree of freedom is sufficient the energy we have to minimize is

\begin{equation} 
E = \int d^{3}x \left[ \frac{1}{2} \dot{\phi}^{2} + \frac{1}{2} |\nabla \phi |^{2} + U(\phi) \right],  
\end{equation} 
while keeping the charge $Q$ constant. It was shown \cite{Kusenko:1997ad} that a useful solution to this extremization problem is offered by the method of Lagrange multipliers. In this approach one minimizes the functional 

\begin{equation} 
\mathcal{E}_{\omega} = E + \omega \left[ Q - \frac{1}{2 i} \int d^{3}x (\phi \partial_{0} \phi^{*} - \phi^{*} \partial_{0} \phi) \right],  
\label{lagrange}
\end{equation} 
where the Lagrange multiplier $\omega$ has the physical interpretation of the mass of a scalar particle inside the $Q$-ball (see Eq. 2).  For a baryoleptonic $Q$-ball one must generalize this procedure to multiple conserved quantum numbers since the baryon and lepton numbers are conserved separately.  In the gauge-mediated scenario the potential is flat out to very large VEV and we can safely take $U(\phi) = M_{s}^{4}$, where $M_{s}$ is the scale of supersymmetry breaking, of order $(1-10) \textsl{TeV}$. Carrying out the minimization procedure on the energy functional yields the following $Q$-ball solution 

\begin{equation} 
\phi(r,t) = \phi_{0} e^{i \omega t} ~\frac{ \sin \omega r}{\omega r}, ~~ r \le R_{Q},\end{equation}
where the $Q$-ball mass is given by Eq. \ref{mass}, the VEV $\phi_{0} \sim M_{s} Q^{1/4}$,  mass per charge $\omega \sim M_{s}/ Q^{1/4}$, and the $Q$-ball radius $R_{Q} \sim \pi /\omega$.

%%%
\section{The Fate of Flat-Direction \boldmath $Q$-balls}
%%%%
In the Affleck-Dine scenario $Q$-balls form along a flat direction which generally consists of both squark and slepton components \cite{Dine:2003ax}.  Although the conditions of $D$- and $F$-flatness are required for the production of $Q$-balls, there is no reason to expect the flat direction state to be the lowest-energy configuration for a $(B+L)$-ball. Moreover, though the baryonic part of these $Q$-balls is constrained by the stability condition, the leptonic part is not.  Thus we might wonder whether the process 

\begin{equation} 
(Q_{B}, Q_{L})_{FD} \longrightarrow (Q_{B},Q_{L} -N)_{GS} + N m_{\nu} , 
\end{equation}
 is kinematically favorable, where the subscripts $FD$ and $GS$ stand for flat-direction and ground state respectively.  We will show that this is indeed the case.
 
 For purposes of illustration let us focus on a toy model which has all the main features of MSSM flat directions. This toy flat direction will have one squark field $q$ and one slepton field $L$. The full potential has the form $U(L,q) = M_{s}^{4} +g^{2} |L^{2} - q^{2}|^{2}$, where $g$ is the coupling constant of some gauge interaction that $q$ and $L$ share. The first term in the potential comes from gauge-mediated supersymmetry breaking and the second term comes from the $D$-terms. As we will see the ordinary condition of $D$-flatness $q=L$ is not upheld for the ground state configuration.  
 
 The first question we may ask about the ground state $Q$-ball is: Do the squark and slepton parts of the $Q$-ball continue to occupy the same volume? That is, does $R_{q} = R_{L}$ continue to hold once we move off the flat direction? Now we generalize Eq. \ref{lagrange} by minimizing with respect to both lepton and baryon numbers separately 
 
 \begin{equation} 
 \mathcal{E}_{\omega, \delta \omega} = (\omega + \delta \omega) Q_{L} + Q_{B} + \int d^{3}x \left[ \frac{1}{2} |\nabla L|^{2} + \frac{1}{2} |\nabla q|^{2} + U(L,q) \right],  \end{equation}
 where $\omega_{q} \equiv \omega$ and $\omega_{L} \equiv \omega + \delta \omega$. Taking $Q_{L} \approx Q_{B}$ in the off-flat direction state and keeping terms $\mathcal{O}(\delta \omega)$ yields 

  \begin{equation} 
  \delta \omega = - \frac{\pi^{2} \omega}{g^{2} Q_{B}}. \end{equation} 
Phenomenologically viable $Q$-balls require $Q_{B} \ge 10^{24}$ \cite{Dine:2003ax}, so that $\delta \omega$ is sufficiently small to justify $\omega_{L} \approx \omega_{q}$. Thus the squark and slepton radii are approximately equal. 

Now we evaluate the mass of the ground-state $Q$-ball 

\begin{equation} M_{f} = \omega_{f} (Q_{f} + Q_{B} ) + \frac{4 \pi M_{s}^{4}}{3 \omega_{f}^{3}} + g^{2} \int d^{3}x |L^{2} - q^{2}|^{2},
\label{finalmass}
 \end{equation} 	
 where the label $f$ is understood to apply to lepton quantities since the baryon number is unaltered. To ensure that the baryon number does not change as we vary $\omega$ we demand that $q_{i}^{2}/ \omega_{i}^{2} = q_{f}^{2}/ \omega_{f}^{2}$. We can write a similar expression relating final and initial lepton amplitudes 
 
 \begin{equation} 
 L_{f} = \left( \frac{Q_{f}}{Q_{i}} \right)^{1/2} \left( \frac{\omega_{f}}{\omega_{i}}\right) L_{i}. \end{equation}
 Using these relations we can carry out the minimization procedure to find
 
 \begin{equation} 
 \omega_{f} = \omega_{i} \frac{1}{\left(1 + N/Q_{f} \right)^{1/4}} \frac{1}{\left(1 + g^{2} \alpha_{4} N^{2}/ \pi^{3}Q_{f} \right)^{1/4}}, \end{equation}
 where $\alpha_{4}$ is an $\mathcal{O}(1)$ constant of integration.
 
 Therefore note that so long as the number of emitted neutrinos is small $Q_{i}$,$Q_{f}$ $\gg N$ then $\omega_{i} \approx \omega_{f}$. We will verify $\it{ex ~post~ facto}$ the validity of the small $N$ assumption. In this approximation the difference in final and initial $Q$-ball masses $\Delta M \equiv M_{i} - M_{f}$
 
 \begin{equation} 
\Delta M = \omega_{i} N - \frac{g^{2} \alpha_{4} \omega_{i}}{\pi^{3}}N^{2},  \end{equation}
 which demonstrates that this decay process is kinematically favorable, $\Delta M \ge N m_{\nu}$, so long as 
 
 \begin{equation} N \le \frac{\pi^{3}}{g^{2} \alpha_{4}} \left(1 - \frac{m_{\nu}}{\omega}\right). \end{equation} 
 Notice that this result is reasonable from a microphysical perspective in the sense that as the mass of a lepton inside the $Q$-ball $\omega$ becomes smaller than the mass outside $m_{\nu}$ the emission turns off. For phenomenologically acceptable values of $Q$ we are always in the limit $\omega \gg m_{\nu}$, implying the upper bound $N \le 480$ for $g^{2} \sim 10^{-1}$. This justifies the validity of the small $N$ approximation and demonstrates that the ground state $Q$-ball is indeed off the flat direction. 
 %%%
 \section{The \boldmath $Q\bar{d}L$ Direction in the MSSM}
 %%%
 We now focus on a more realistic scenario in which a $Q$-ball forms along the $Q_{1} \bar{d_{2}} L_{1}$ direction, where subscripts are generation indices. This direction can be parameterized as 
 
\begin{equation} Q_{1} = \frac{1}{\sqrt{3}} \left( \begin{array}{c}
\phi  \\
0 \end{array}  \right), ~~~
L_{1} = \frac{1}{\sqrt{3}} \left(\begin{array}{c}
0 \\
\phi \end{array} \right), ~~~
\bar{d_{2}} =\frac{1}{\sqrt{3}} \phi. 
\end{equation}
 The complete potential for this direction is $U = M_{s}^{4} + U_{D} + U_{NR}$, where the first term comes from supersymmetry breaking, the second term are the $D$-terms and the last term comes from non-renormalizable operators.  The $SU(2) \times U(1)$ $D$-terms are

\begin{equation} \label{true} U_{D}   =  \frac{g^{2}}{8}\left( |Q_{1}|^{2} - |L_{1}|^{2}\right)^{2} +
\frac{g'^{2}}{72}\! \! \left ( |Q_{1}|^{2} - 3|L_{1}|^{2} + 2
|\bar{d_{2}}|^{2}\right )^{2} , \end{equation}
where $g$ and $g'$ are the $SU(2)$ and $U(1)$ couplings respectively. Of course $SU(3)$ $D$-terms exist as well but they are unimportant since the two squark fields have the same amplitudes and same $\omega$ value. The primary difference between this case and our toy example however are the non-renormalizable terms which for this direction have the form

\begin{equation} U_{NR} (\phi) = \frac{|\lambda|^{2}}{M^{2}} |\phi|^{6}, \end{equation}
 where $M$ is a large mass scale such as the GUT or Planck scale. 
 
 Since both the squark and slepton fields vary from their flat-direction values, they both contribute to the final $Q$-ball mass 
 
 \begin{equation} (\Delta M)_{NR} = \frac{|\lambda|^{2}}{M^{2}} \int d^{3}x \left(|L_{i}|^{6} - |L_{f}|^{6}\right) + \left(|q_{i}|^{6} - |q_{f}|^{6}\right).
 \end{equation} 
 Operating in the $N \ll Q_{i}$ we obtain $(\Delta M )_{NR} = - |\lambda|^{2} \alpha_{6} \omega_{i}^{3} Q_{i}^{2} N / 24 \pi^{6} M^{2}$. This gives an important modification to the emission 
 
 \begin{equation} N \le \frac{\pi^{3}}{\lambda \alpha_{4}} \left( 1- \frac{m_{\nu}}{\omega} - \frac{ |\lambda|^{2} \alpha_{6} \omega_{i}^{2} Q_{i}^{2}}{24 \pi^{6} M^{2}}\right). \end{equation} 
 Assuming $\lambda \sim 10^{-1}$ and $M \sim M_{pl}$, we see that for $Q_{i} \ge 10^{26}$ the flat-direction state becomes the true ground state, thereby shutting off any neutrino emission. 
 \section{Competing Decay Channels}
In addition to neutrinos the $Q$-balls can also emit electrons. The number of emitted electrons will be smaller than the number of neutrinos since they are both charged and more massive. Now we allow the process 
\begin{equation} 
(Q_{B}, Q_{L})^{0}_{FD} \longrightarrow (Q_{B},Q_{L} -N_{e} - N_{\nu})^{+}_{GS} + N_{\nu} m_{\nu} + N_{e} m_{e} , 
\end{equation}
where the subscripts on the $Q$-ball states indicate the electric charge sign of the $Q$-ball.  For nonzero $N_{e}$ this represents a novel form of electric charge acquisition. We can generalize Eq. \ref{finalmass} by including the Coulomb energy of the now electrically-charged $Q$-ball 
\begin{equation}  
M_{f} = \omega_{f} (Q_{f}+Q_{B}) + \frac{4\pi U_{0}}{3 \omega_{f}^{3}} +\frac{3 e^{2}N_{e}^{2}
\omega_{f}}{20 \pi^{2}} + \lambda \int d^{3}x |L^{2} - q^{2}|^{2}.  
\end{equation}
With both leptonic decay channels open the condition for decay $\Delta M \ge \sum_{i} N_{i} m_{i}$ becomes

\begin{equation}
 \label{equal} \omega  \left[(N_{e} + N_{\nu}) - \frac{\alpha_{4}\lambda (N_{\nu}+ N_{e})^{4}}{
\pi^{3}} -\frac{3e^{2}N_{e}^{2}}{20  \pi^{2}} \right] \ge N_{e} m_{e} + N_{\nu}m_{\nu}. \end{equation}
Rather than setting a hard limit on the number of emitted leptons, the decay condition now carves out a kinematically allowed region in the $N_{e} - N_{\nu}$ plane. To determine how a given $Q$-ball actually evolves we now must examine decay rates. 

\begin{figure}
\begin{center}
\includegraphics[scale=0.7]{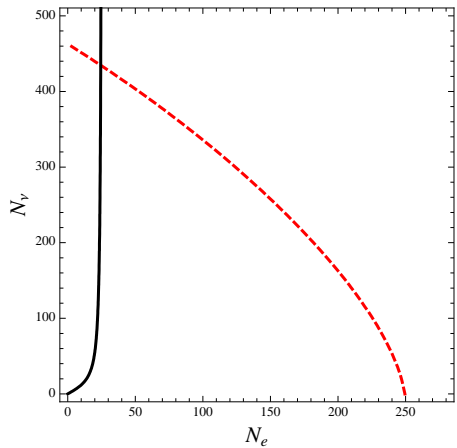}
\caption{The region below the dashed curve is the kinematically allowed regions of decay from a flat-direction $Q$-ball with $Q_{i} = 10^{24}$. The solid curve shows actual dynamical trajectory in the $N_{\nu}-N_{e}$ plane as determined by the leptonic decay rates. }
\end{center}
\label{figure1}
\end{figure}

The decay rates depend on the mass and electric charge of a given lepton. More specifically, for sufficiently large $Q$-balls the rate of decay into fermions is limited by the number of fermions that can form inside and cross the $Q$-surface per unit time \cite{Cohen:1986ct}. For arbitrary mass and charge this rate is 

\begin{equation} \frac{dQ}{dt} \le \frac{A}{24 \pi^{2}} k_{max}^{3}, \end{equation} 
where $k_{max}$ is defined implicitly by 

\begin{equation} 
\omega = \sqrt{k_{max}^{2}+m^{2}} + \frac{3 e^{2}N_{e}^{2} \omega}{20 \pi^{2}}, \end{equation}
for a $Q$-ball of electric charge $N_{e}$. We can use these decay rates to plot the trajectory of the $Q$-ball as it evolves in $N_{e}$-$N_{\nu}$ plane. This is done for a representative $Q$-ball charge $Q_{i} = 10^{24}$ in Fig. \ref{figure1} The intersection of the energetically allowed boundary with the dynamic trajectory gives the maximal electric charge a $Q$-ball can acquire as it evolves to the ground-state. This gives a maximal electric charge $N_{e} \le 25$ for this $Q$-ball. 

%%%
\section{Experimental Constraints}
%%%
In the previous section we outlined a novel mechanism by which a $Q$-ball can acquire electric charge. This is phenomenologically interesting because the electric charge has dramatic consequences for their experimental signatures \cite{Kusenko:1997vp}. Neutral $Q$-ball have a spectacular signature in which they transform incoming protons into antiprotons with an $\mathcal{O}(1)$ probability \cite{Kusenko:2004yw}. 
However for electrically charged $Q$-balls this process is Coloumb suppressed and their main interaction with matter is electromagnetic in nature.

The experimental constraints for baryoleptonic $Q$-balls are altered now in light of the previous sections. In general the constraints on electrically charged $Q$-balls are a few orders of magnitude more stringent requiring $Q \ge 10^{30}$ (for $Z_{Q} \sim 10$), as compared to their neutral counterparts which need $Q \ge 10^{24}$ (taking $M_{s} \sim 10^{3} \textsl{GeV}$) \cite{Ambrosio:1999gj, Arafune:2000yv, Takenaga:2006nr}. 

 \section{Conclusions}
$Q$-balls are attractive dark matter candidates that can naturally be produced in the same process that created the baryon asymmetry. Although they are created in a flat-direction state, we have shown that those of the baryoleptonic variety decay into the ground-state configuration by emitting some number of leptons. The emission of electrons allows the $Q$-ball to gain electric charge in a novel process. This changes experimental limits on baryoleptonic dark matter $Q$-balls.

\acknowledgements The author wishes to thank Alexander Kusenko for suggesting this work.

\bibliography{qball}
\bibliographystyle{apsrev}

\end{document}